\documentclass[10pt, conference]{IEEEtran}

\IEEEoverridecommandlockouts
\usepackage[utf8]{inputenc}
\usepackage{color}
\usepackage{booktabs}
\usepackage{url}
\usepackage{fancybox}
\usepackage{xcolor}
\usepackage{multirow}
\usepackage{float}
\usepackage{array,graphicx}
\usepackage{subcaption}
\usepackage[flushleft]{threeparttable}
\usepackage[most]{tcolorbox}
\usepackage{xspace}
\usepackage{listings}
\usepackage{pifont}
\usepackage{tikz}
\usepackage{balance}

\definecolor{purplish}{HTML}{D8D0E3}
\definecolor{purplishlight}{HTML}{EBE7F1}
\definecolor{purplishdark}{HTML}{168591}

\newcommand{\FunnelBot} {\textsc{FunnelBot}\xspace}

\newcommand*\circled[1]{\tikz[baseline=(char.base)]{
            \node[shape=circle,draw,inner sep=2pt] (char) {#1};}}

\newtcolorbox[auto counter,number within=section]{rqbox}[2]{
    label=#2, nameref=#1,
    title=\small{#1}, 
    enhanced,
    attach boxed title to top left={yshift=-6pt, xshift=8pt},
    boxed title style={size=small,boxsep=1pt},
    colframe=purplishdark,colback=white,colbacktitle=purplishdark,
    boxsep=2pt,left=2pt,right=2pt,top=6pt,bottom=2pt,middle=2pt
}

\newcommand{\takeayaone}[2][]{
    \begin{rqbox}{\textbf{Takeaway message \#1}}{#2}
        #1
    \end{rqbox}
}

\newcommand{\takeayatwo}[2][]{
    \begin{rqbox}{\textbf{Takeaway message \#2}}{#2}
        #1
    \end{rqbox}
}

\begin{document}

\title{Together or Apart? Investigating a mediator bot to aggregate bot's comments on pull requests}

\author{\IEEEauthorblockN{
Eric Ribeiro\IEEEauthorrefmark{1},
Ronan Nascimento\IEEEauthorrefmark{1},
Igor Steinmacher\IEEEauthorrefmark{2}, 
Laerte Xavier\IEEEauthorrefmark{1},\\
Marco Gerosa\IEEEauthorrefmark{2},
Hugo de Paula\IEEEauthorrefmark{1}, and
Mairieli Wessel\IEEEauthorrefmark{3}}
\IEEEauthorblockA{\IEEEauthorrefmark{1}Pontifical Catholic University of Minas Gerais, Brazil}
\IEEEauthorblockA{\IEEEauthorrefmark{2}Northern Arizona University, USA}
\IEEEauthorblockA{\IEEEauthorrefmark{3}Radboud University, The Netherlands}
}

\maketitle

\begin{abstract}
Software bots connect users and tools, streamlining the pull request review process in social coding platforms. However, bots can introduce information overload into developers' communication. Information overload is especially problematic for newcomers, who are still exploring the project and may feel overwhelmed by the number of messages. Inspired by the literature of other domains, we designed and evaluated \FunnelBot, a bot that acts as a mediator between developers and other bots in the repository. We conducted a within-subject study with 25 newcomers to capture their perceptions and preferences. Our results provide insights for bot developers who want to mitigate noise and create bots for supporting newcomers, laying a foundation for designing better bots. 

\end{abstract}


\section{Introduction}

On social coding platforms such as GitHub, developers are often overwhelmed by bot pull request notifications, which interrupt their workflow~\cite{Wessel2021CSCW}. As pointed out by Wessel \textit{et al.}~\cite{Wessel2021CSCW}, as bots have become new voices in developers' conversations, they may overburden developers who already suffer from information overload when communicating online~\cite{nematzadeh2016information}. This problem is especially relevant for newcomers, who require special support during the onboarding process due to the barriers they face~\cite{steinmacher2015social}. Newcomers can perceive bots' complex answers as discouraging, as bots often provide a long list of contribution feedback items (e.g., style guidelines, failed tests) rather than supportive assistance.

Developing ways to deal with the information overload bots cause is critical for the future of bots in software development. In other domains, researchers have proposed meta-bots to integrate and moderate the interactions of multiple bots that cannot be adapted~\cite{sadeddin2007online}. In the scope of social coding platforms, Wessel \textit{et al.}~\cite{Wessel2022ICSE} envisioned a mediator bot as a promising approach to mitigate the information overload from existing GitHub bots. However, to date, the design of such a mediator bot in a social coding platform has not been investigated. In this paper, we close this gap by understanding how newcomers consume the information given by \FunnelBot, a mediator bot that categorizes and organizes the information from multi-bots into a single comment in the pull request interface.

Understanding newcomers' preferences is important to effectively design bot messages and reduce noise. In this paper, we evaluate how grouping and categorizing bots' outputs compares to the traditional GitHub approach (i.e., multi-bots commenting on the pull request). Our study focuses on answering the following research question: \textit{How do newcomers perceive the multi-bot aggregated information?}

%\rqperception{rqn:perception}

To answer this research question, we conducted a within-subject study with 25 participants, including undergraduate and graduate students who may or may not have previous experience with bots and open-source software development, but are newcomers to the investigated project. Participants interacted with each information presentation approach (\FunnelBot vs. multi-bots) individually and comparatively. Finally, we compared the perceived usefulness of the approaches by applying surveys and qualitatively analyzing the participants' feedback. 

We found that newcomers perceive the information provided by multi-bots and \FunnelBot as appropriate, clear, easy to understand, and useful. However, they have diverse opinions regarding the amount of information provided. Also, participants reported that aggregating bot comments helped developers find the appropriate information. With these findings, we take a step towards understanding newcomers' preferences about bots' feedback on pull requests.  We made the data publicity available for replication purposes.\footnote{\url{https://zenodo.org/record/5596321}} 

\section{Related Work}
\label{sec:related-works}

Despite the widespread adoption of bots on social coding platforms, the interaction between bots and humans still presents challenges~\cite{Wessel2018,Wessel2021CSCW}. Analyzing the \textit{tool-recommender-bot}, Brown and Parnin~\cite{Brown2019} report that bots still need to overcome problems such as notification overload. Peng and Ma~\cite{Peng2019} studied how developers perceive and work with \textit{mention bot} and concluded that it does not meet users' diverse needs. Results show that developers are bothered by frequent review notifications when dealing with a heavy workload. These results are in line with the study conducted by Wessel \textit{et al.}~\cite{Wessel2021CSCW}, which indicates that noise is a central problem. Noise affects human communication and the development workflow by overwhelming and distracting developers. 
Wessel et al.~\cite{Wessel2022ICSE} have proposed the concept of mediator bot, which aggregates and summarizes the information coming from several bots, to alleviate the information overload. Complementing the previous literature, we focus on understanding how to adjust the interaction of existing bots to improve the experience of newcomers.

\section{Research Design}

\subsection{\FunnelBot prototype}

\FunnelBot was inspired by the meta-bot proposed by Wessel \textit{et al.}~\cite{Wessel2022ICSE}. 
To build this idea, we focused on mitigating communication noise by preventing bots from interacting directly on pull requests. \FunnelBot provides an interface that allows each project to configure bots' restrictions. Therefore, once a new pull request is opened, \FunnelBot aggregates only the responses of bots allowed to interact with that pull request. 

\begin{figure}[htb]
    \centering
    \includegraphics[width=\linewidth]{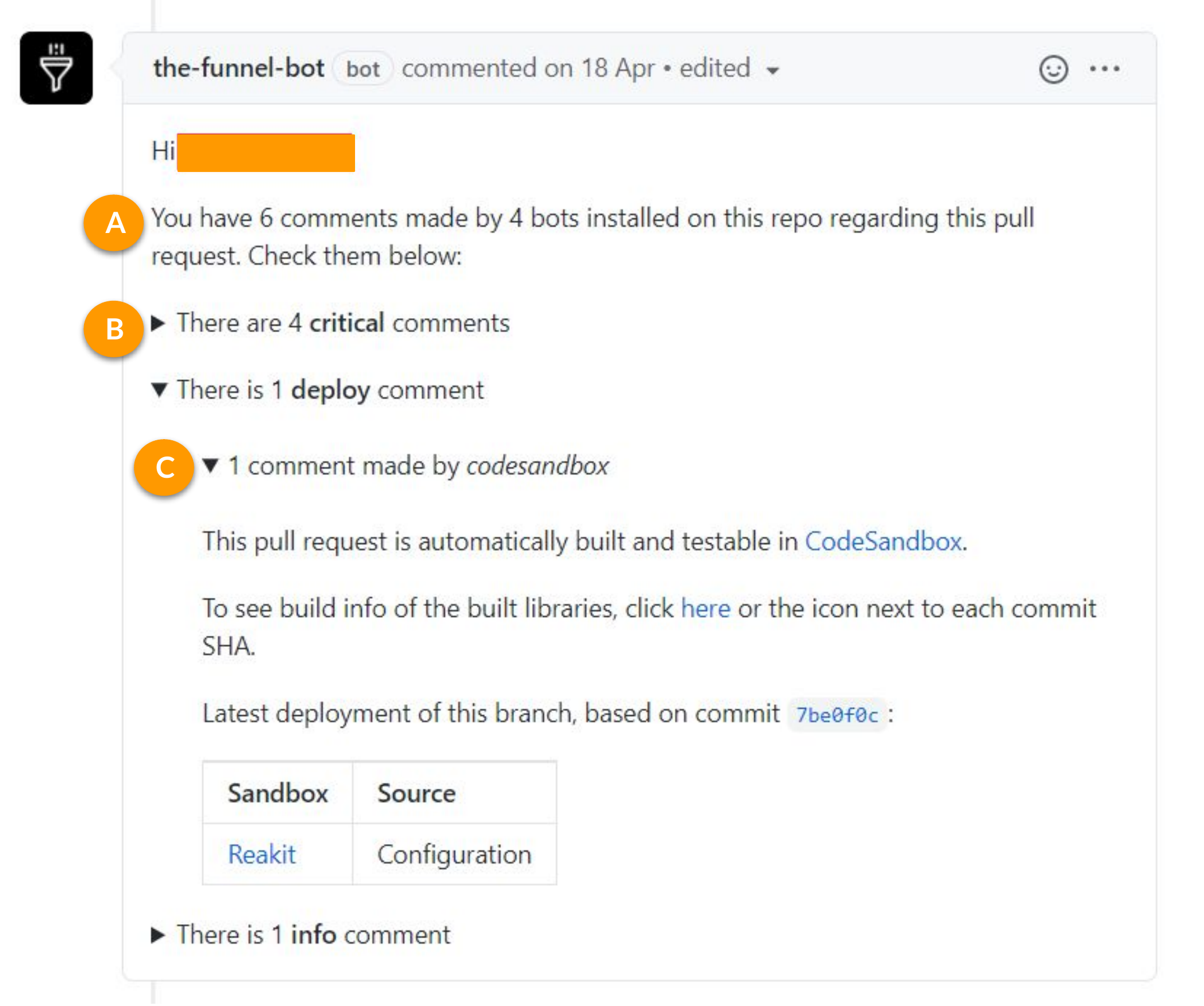}
    \caption{Example of pull request comment posted by \FunnelBot}
    \label{fig:metabot-comentario-completo}
\end{figure}

Figure \ref{fig:metabot-comentario-completo} shows an example of pull request comment posted by our bot. The comment shows an introductory message (\circled{A} in Figure~\ref{fig:metabot-comentario-completo}), a list with all groups of bot messages collapsed (\circled{B} in Figure~\ref{fig:metabot-comentario-completo}), and one expanded example where we can see the \texttt{CodesandBox} comment (\circled{C} in Figure~\ref{fig:metabot-comentario-completo}). We implemented \FunnelBot using Probot,\footnote{https://probot.github.io/docs/} a framework developed by GitHub. We made the description of the \FunnelBot's architecture and implementation publicity available in our replication package.

\subsection{\FunnelBot evaluation}

To showcase \FunnelBot to our participants in a real-world scenario, we created a copy of the open-source Ariakit\footnote{\url{https://github.com/ariakit/ariakit}} project, since it uses four bots to support the pull request review process. Each bot is responsible for a different task: \texttt{reakit-bot}, \texttt{codesandbox}, \texttt{compressed-size-action}, and \texttt{codecov}. All bots report their outputs using comments on the pull request. Since \texttt{reakit-bot} is a closed-source project-specific bot, and not accessible in the GitHub Marketplace, we could not use it in the experiment. Therefore, we replace the \texttt{reakit-bot} with two well-known and highly utilized bots: \texttt{request-info} bot and \texttt{TODO} bot.

\noindent\textbf{\textit{Sandbox sessions}} --
We conducted sandbox sessions with one Ph.D. student and four developers who are novices contributing to open source projects on GitHub. We validated the script and confirmed whether the session would fit in a 30-minute time slot. The participants suggested a few minor adjustments, which we incorporated into the instruments. The data collected during these sandbox sessions were discarded.

\noindent\textbf{\textit{Participants recruitment}} --
We used convenience sampling to recruit participants for our experiment sessions, inviting students from the institution where one of the authors is a faculty member. The study was not part of any course, all the participants were volunteers, and all signed a consent before their sessions.
In total, 22 undergraduate and 3 graduate students were recruited. Their experience contributing to open-source ranged between none and experienced (with low experience on average). Finally, their experience with GitHub bots varied between none and experienced (with no experience on average). Six participants reported having some experience with one or more GitHub bots (e.g.,  \texttt{dependabot}). All other participants (19) reported no previous interaction with bots.

\noindent\textbf{\textit{Experimental sessions}} -- 
We conducted a series of synchronous within-subject sessions via video calls since it enables us to guide participants throughout the experiment. Participants received email instructions for the experiment, a survey with demographic questions, and an invitation to the video call. We recorded the audio and took notes of all sessions. The sessions lasted on average 45 minutes.

We started the session with a short explanation about the research objectives and guidelines. We explained that they would play the role of a new contributor to the Ariakit project who has just submitted a pull request. The first scenario presented to participants (\textbf{\textit{Scenario 1}}) is based on Ariakit's pull request \#796.\footnote{\url{https://github.com/reakit/reakit/pull/796}} All comments from humans and bots have been replicated in two distinct versions of pull request \#796  (\textit{S1-FunnelBot} and \textit{S1-Multibots}) with the distinction that one (\textit{S1-Funnel}) has its bots comments aggregated by \FunnelBot. Thirteen participants (Group 1) interacted with pull request \textit{S1-Multibots}; while twelve participants (Group 2) interacted with pull request \textit{S1-FunnelBot}. Participants were asked to use the information presented in the pull request to identify its current status as well as the next steps for the pull request to be accepted by the maintainers. After the participant has completed the pull request analysis, we directed them to a survey to collect their perceptions.

At this point, each of the participants had just interacted with a single approach (multibots or \FunnelBot). We then moved to \textbf{\textit{Scenario 2}}, which was created based on the pull request \#828.\footnote{\url{https://github.com/reakit/reakit/pull/828}} Participants who started with multibots approach in Scenario 1 (S1-Multibots), interacted with \FunnelBot (S2-FunnelBot) approach in Scenario 2 and vice-versa. The group of participants who analyzed the pull requests \textit{S1-Multibots} followed by \textit{S2-FunnelBot} is referred to as Group 1 (G1), while the group who started with \textit{S1-FunnelBot} followed by \textit{S2-Multibots} is referred to as Group 2 (G2). Once again, we guided the participants to look at the pull request to understand its current status and pending tasks. As for Scenario 1, we redirected participants to a survey after completing the pull request analysis. This final survey aimed at comparing the two approaches: \FunnelBot vs. multi-bots.

\noindent\textbf{\textit{Data Analysis}} --
To statistically compare the distributions of the close-ended questions between group 1 and group 2, we used the non-parametric Mann-Whitney-Wilcoxon test~\cite{wilks2011statistical}. In this context, the null hypothesis ($H_0$) is that the distributions of group 1 and 2 answers are the same, and the alternative hypothesis ($H_1$) is that these distributions differ. We also use Cliff's Delta~\cite{romano2006appropriate} to quantify the difference between these groups of observations beyond $p$-value interpretation. According to Romano et al.~\cite{romano2006appropriate}, the magnitude of delta ($|\delta|$) is assessed using the following thresholds: $|\delta| <$ 0.147 ``negligible'', $|\delta| <$ 0.33 ``small'', $|\delta| <$ 0.474 ``medium'', otherwise ``large.''

We used a card sorting approach~\cite{zimmermann2016card} to analyze the answers to the open-ended questions. The first two authors conducted card sorting in two steps. In the first step, each researcher independently analyzed the answers (cards) and applied codes to each answer, sorting them into meaningful groups. Discussion meetings followed this step until reaching a consensus on each item's code name and categorization. The answers were sorted into high-level groups at the end of this process. In the second step, the researchers analyzed the categories, aiming to refine the classification and group related codes into more significant, higher-level categories and themes. We used \emph{open} card sorting, meaning we had no predefined codes or groups; the codes emerged and evolved during the analysis process. 

\section{Results}

\subsection{Newcomers' perceptions on \FunnelBot vs. multi-bots}

\begin{figure}[!htbp]
    \scriptsize
    \centering
    \includegraphics[width=1\linewidth]{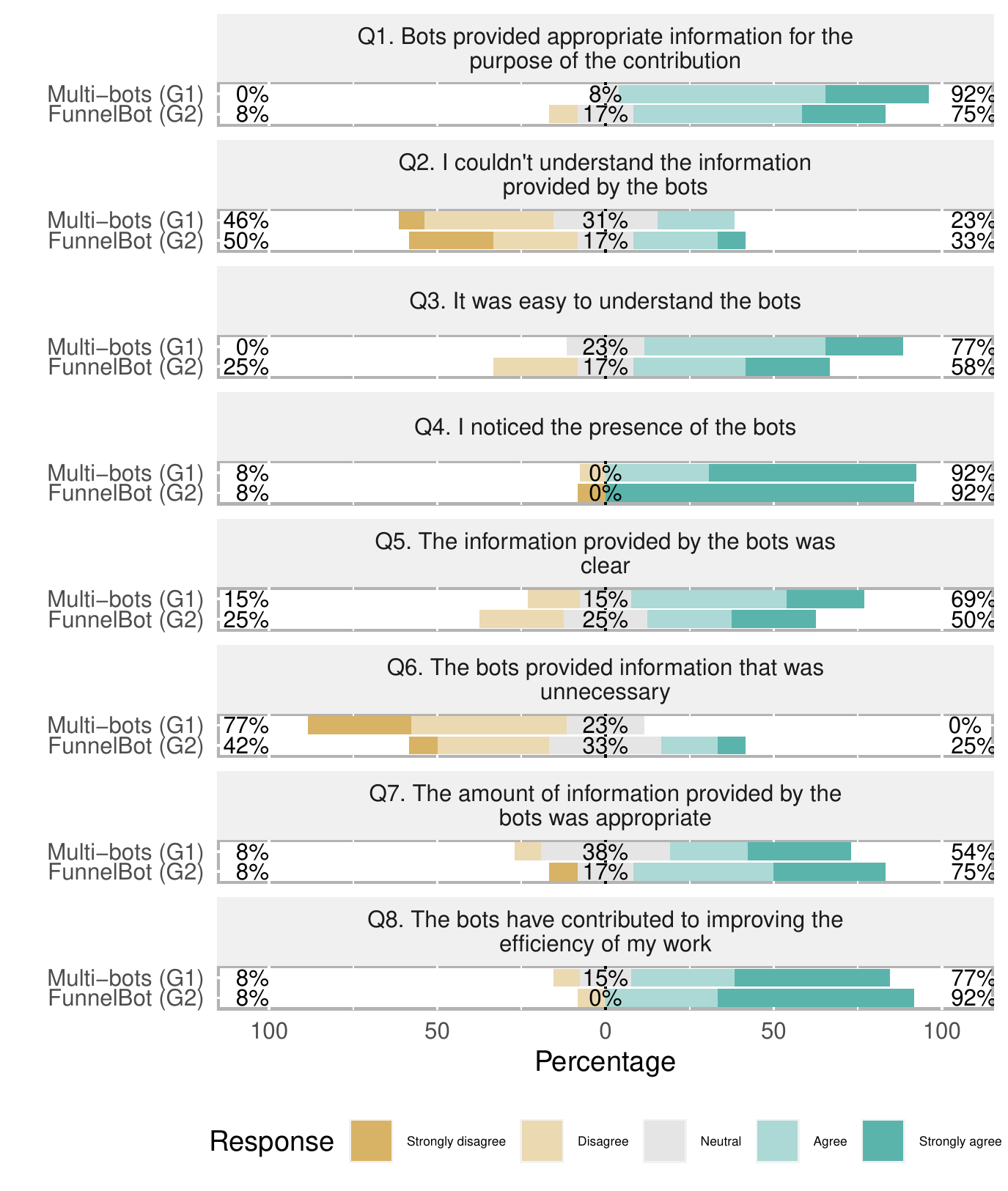}
    \caption{Participants' perceptions about the approaches}
    \label{fig:Perception}
\end{figure}

In this section, we focused on the results obtained after the participants concluded Scenario 1. Our goal was to compare the multi-bots \textit{versus} \FunnelBot approach (see Figure~\ref{fig:Perception}).

% \begin{table}[htb]
%     \caption{Results of Mann-U test Multi-bots vs. \FunnelBot}
%     \label{tab:perceptions}
%     \centering
%     \begin{tabular}{r|l}
%         \toprule
%         Likert Item & p-value \\
%         \midrule
%         Q1 & 0.412  \\
%         Q2 & 0.910  \\
%         Q3 & 0.420  \\
%         Q4 & 0.135  \\
%         Q5 & 0.571  \\
%         Q6 & 0.037* \\
%         Q7 & 0.569  \\
%         Q8 & 0.472  \\
%         \bottomrule
%     \end{tabular}
% \end{table}

Most participants in group 1 (G1) and group 2 (G2) (92\% and 75\%, respectively) claimed that bots offered appropriate information (Q1) for them to decide the next steps (i.e., if any modification is needed in the pull request before merging). We did not find a statistically significant difference ($p$-value $0.412$) between G1 (that interacted with the multi-bots approach) and G2 (that interacted with \FunnelBot).

Regarding how clear the bots' comments are (Q2 and Q5), the response was also very similar between the two groups. For Q2, 46\% of G1 participants disagreed or strongly disagreed that they could not understand the information, while 50\% of G2 disagreed ($p$-value $0.910$). For Q5, both groups agreed (69\% and 50\%, respectively) that the information provided by the multi-bots (or \FunnelBot) was clear. We also did not find a statistically significant difference ($p$-value $0.571$) between G1 and G2.

Regarding the amount of information offered by the bots (Q6), the two groups had diverse opinions ($p$-value $0.037$, $\delta=0.474359$). The majority (77\%) of G1 participants disagreed that the information offered by the bots was unnecessary, and the remaining (23\%) were neutral. In G2, 42\% of the participants disagreed that the information was redundant, while 33\% were neutral, and 25\% answered that the information was unnecessary for them.

Participants also mentioned that it was easy to understand the information provided by both approaches (77\% and 58\%, for G1 and G2, respectively). In Q7, participants also mentioned that the amount of the information provided by the bots was appropriate (54\% and 75\%, for G1 and G2, respectively).

Regarding the perception of bots' presence in the pull request (Q4), the majority of participants in both groups agreed or strongly agreed that they perceived the presence of multiple bots (or \FunnelBot) in the pull request (92\% for both groups, with $p$-value $0.135$). Regarding the usefulness of bot comments, both groups agreed (77\% of G1 and 92\% of G2) that bots' information has improved efficiency (Q8, $p$-value $0.472$) when performing tasks.

\takeayaone[
Newcomers perceive the information provided by the multi-bots and \FunnelBot approaches as appropriate, clear, easy to understand, and useful.
]{}

\subsection{Comparison of newcomers' preferences}

\begin{figure}[!htbp]
    \scriptsize
    \centering
    \includegraphics[width=1\linewidth]{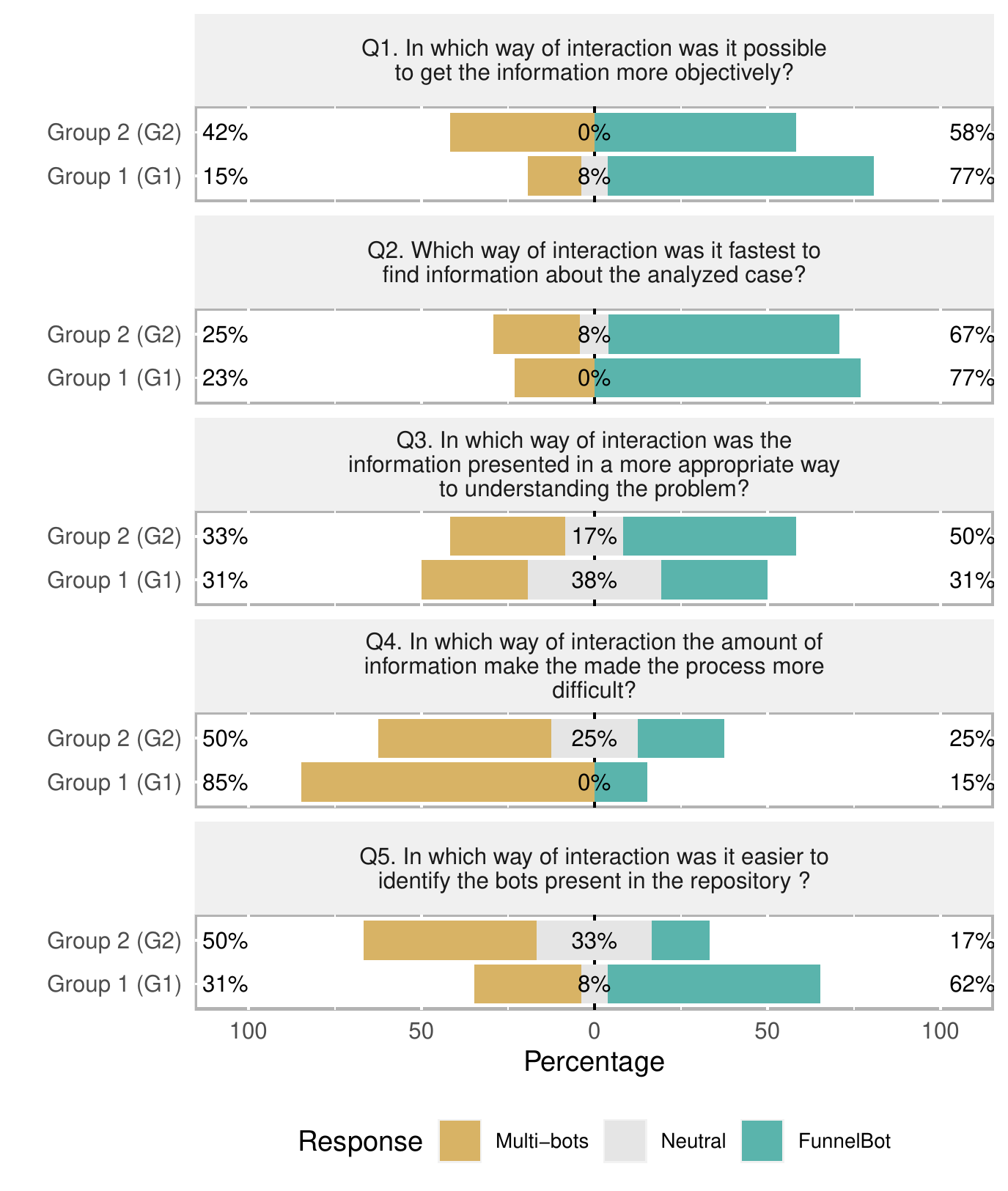}
    \caption{\FunnelBot vs multi-bots comparison}.
    \label{fig:Comparison}
\end{figure}

We also sent our participants a questionnaire to compare both approaches (see Figure~\ref{fig:Comparison}).

Most participants in G1 and G2 (77\% and 58\%, respectively) identified the \FunnelBot approach as facilitating quicker location of information. Most participants (77\% in G1 and 67\% in G2) considered \FunnelBot approach the fastest way to find information in the analyzed scenario. A G1 participant described a situation to explain his preference for \FunnelBot: ``\textit{if I am responsible for deploying the application, I will know accurately and quickly where to look for information regarding my responsibilities in that repository [...]}.'' A G2 participant reported that, when using \FunnelBot, the ``\textit{information [is] aggregated in a comment containing all the relevant information generated by the bot},'' which facilitates access to such information.

Participants had different opinions about which approach made it easier to understand the provided information. In G1, participants reported that the organization of the information had not affected their understanding. One participant mentioned that when they know the category of the problem reported by the bot (e.g., deploy, critical), they prefer \FunnelBot because of the comment categorization. However, if they need to analyze the entire pull request, they prefer the various bots, as the information is presented along with the pull request timeline. In G2, 50\% of the participants found \FunnelBot's approach as the easiest for understanding the problem reported by the bots. One participant mentioned that \FunnelBot ``\textit{has the same benefits as the various bots; however, it aggregates the information, which helps to understand the information}''.

In both G1 and G2, the participants (85\% and 50\%, respectively) said that the interaction with several bots made it challenging to search for information. Furthermore, most G1 participants (62\%) stated that it was easier to identify the bots present in the repository when interacting with \FunnelBot. A participant from this group reported that the way \FunnelBot comment was structured drew attention and highlighted the bots commenting on the pull request. In contrast, a more significant number of G2 participants (50\%) stated that it was easier to identify the bots present in the approach with several bots.

We also asked participants about the benefits of the multi-bots approach and the benefits of \FunnelBot. Concerning the benefits of \FunnelBot, the most frequent answer refers to its ability to \textbf{facilitate finding important information} (8 and 4 mentions, in G1 and G2, respectively). The participants explained this by mentioning that \FunnelBot \textbf{reduces information overload} (3 mentions in G1), \textbf{categorizes bot comments} (5 and 4 mentions, in G1 and G2 respectively) and \textbf{aggregates comments} (5 mentions in both groups). One participant stated that ``\textit{for people used to contributing to open-source projects daily, [\FunnelBot] keeps the information aggregated into a single comment, giving more visibility to developers' comments.}''

\takeayatwo[
    Although some people reported they preferred the multi-bot, by analyzing the open-ended questions we noticed that the information was not useful and, sometimes, disregarded by the developers---showing that the \FunnelBot better organizes the information.
]{}

\section{Discussion}

Developers are constantly relying on bots to stay productive~\cite{Storey2016}. However, the information overload that some bots introduce can have the opposite effect~\cite{Wessel2020}. To address this issue, \FunnelBot introduces a novel approach concerning information organization on pull requests, making it an alternative to the existing strategy used on GitHub. Our study contributes with an empirical analysis comparing the current bots' information presentation style in GitHub with the approach proposed by \FunnelBot.

After completing the second scenario and having contact with both forms of information presentation, both groups of participants agreed that the multi-bot approach makes it difficult to find information. Nine participants pointed to the information overload derived from the comments from the multiple bots as the cause of this difficulty. These participants identified reducing information overload as a positive point of \FunnelBot. Therefore, our results demonstrate \FunnelBot as a viable approach to reduce the cognitive load introduced by bots.

\noindent\textbf{Implications.}
It is known that developers with different profiles and backgrounds have different expectations and preferences for interacting with bots~\cite{Wessel2018,Wessel2021CSCW}. Additional effort is still necessary to investigate how the strategies of aggregating bots' information might influence the way developers interpret the bot comments' content. How developers think, perceive, and remember information (i.e., their cognitive style) is likely to affect how they respond to bot messages and learn from them~\cite{gendermag}.

The preliminary implementation of \FunnelBot revealed some limitations imposed by the GitHub platform that restrict the design of bots. Wessel \textit{et al.}~\cite{Wessel2021CSCW} already mentioned some examples of those technical challenges in their hierarchical categorization of bot problems. In short, the platform restrictions might limit both the extent of bot actions and the way bots are allowed to communicate. It is essential to provide a more flexible way for bots to interact on the platform. We prototyped the strategy of aggregating bot events by designing a new bot that interacts in the pull request interface; this idea can be leveraged to reshape the interface and better accommodate bot interactions.

\section{Limitations}

We are aware that each bot and project has its singularities and that the open-source universe is expansive---our strategy to continue collecting data until reaching information saturation aimed to alleviate this issue. Moreover, our sample was composed of students. Although they are newcomers~\cite{Steinmacher.Conte.ea_2016}, we acknowledge that additional research is necessary to consider the perspective of practitioners experienced with bots.
The introduction of categorization bias may have occurred during the analysis of the open-ended questions. Therefore, the categorization process was carried out in pairs with a careful discussion among the authors regarding the categories. We also tried to order the questions based on the natural sequence of actions to help respondents understand the questions' context.

\section{Conclusion}

We evaluated a mediator bot, called \FunnelBot, by comparing it with the current approach used by GitHub. The evaluation of \FunnelBot showed that aggregating bot comments was helpful to facilitate developers in finding and processing the information from bots, reducing the overload. Although some participants reported a good experience when using the multi-bot approach, their interaction showed that they misunderstood the content of the bots' comments. Social coding platforms can implement a strategy of aggregating and classifying comments to mitigate some interaction problems in pull requests introduced by bots, and bot developers can redesign how they present information.

\balance
\bibliographystyle{IEEEtran}
\bibliography{references}

\end{document}